\documentclass[useAMS,usenatbib]{mn2e}
\usepackage{epsfig}

\title[Evolution of M82-like starburst
galaxy winds ]{Evolution of  M82-like starburst 
 winds revisited: 3D radiative cooling hydrodynamical simulations}
\author[C. Melioli, E. M. de Gouveia Dal Pino, F. G. Geraissate,]
{C. Melioli,$^{1}$\thanks{E-mail:
cmelioli@astro.iag.usp.br; dalpino@astro.iag.usp.br; 
geraissate@astro.iag.usp.br} 
E. M. de Gouveia Dal Pino$^{1}$ and F. G. Geraissate$^{1}$ \\
$^{1}$IAG-Universidade de Sao Paulo, Rua do Matao 1226, Sao Paulo, SP, Brazil}

\begin{document}

\date{Accepted 18 Jan 2012}

\pagerange{\pageref{firstpage}--\pageref{lastpage}} \pubyear{2012}

\maketitle

\label{firstpage}

\begin{abstract}

In this study we present three-dimensional radiative cooling hydrodynamical 
simulations of galactic winds generated particularly in M82-like starburst 
galaxies.
We have considered intermittent winds induced by SNe explosions within super 
star clusters randomly distributed (in space and time) in the central region of 
the galaxy (within a radius R = 150 pc) and were able to reproduce the 
observed M82 wind conditions with its complex morphological outflow structure.
We have found that the environmental conditions in the disk in nearly recent 
past are crucial to determine whether the wind will develop a large scale rich 
filamentary structure, as in M82 wind, or not.  If a sufficiently large 
number of super stellar clusters is build up in a starburst mainly over a 
period of a few million years, then the simulations reproduce the multi-phase 
gas observed in M82-like winds, i.e., with filaments of sizes  about 20 
to 300 pc, velocities of $\sim$ 200 - 500 km/s, densities in the range of 
$10^{-1}$ - 10 cm$^{-3}$, embedded in a hot low density gas with a density 
smaller than $10^{-2}$ cm$^{-3}$ and a velocity of $\sim$ 2000 km s$^{-1}$. 
Otherwise, a "superbubble-like" wind develops, with very poor or no cold 
filamentary structures.  
Also, the numerical evolution of the SN ejecta have allowed us to obtain the 
abundance distribution over  the first $\sim$ 3 kpc extension of the wind and 
we have found that the SNe explosions change significantly the metallicity 
only of the hot, low-density wind component for which we obtained  abundances 
$\sim$ 5 to 10 $Z_{\odot}$ in fair consistency with the observations. 
Moreover, we have found that the SN-driven wind transports to outside the 
disk large amounts of energy, momentum and gas, but the more massive 
high-density component reaches only intermediate altitudes smaller than 1.5 
kpc. Therefore, no significant amounts of gas mass are lost to the IGM and the 
mass evolution of the galaxy is not much affected  by the starburst events 
occurring in the nuclear region.

\end{abstract}


\section{Introduction} 
 \label{sec:introduction}

The evolution of the interstellar medium (ISM) at galactic scales
depends on the stellar feedback at the larger scales, and vice-versa. 
In fact, massive stars exert  a profound influence over the baryonic 
component of the galaxy. At the same time, the collective effect of supernova 
(SN) explosions, stellar winds and radiation pressure may drive galactic winds 
(GWs; see, e.g., \citealt{mat71}, \citealt{heck90}, \citealt{veil05}).
GWs influence the multi-phase conditions of the ISM \citep{mckee95}  as well 
as the dynamical and chemical evolution of the galaxy and the thermodynamics 
and enrichment of the intergalactic medium (IGM; see, e.g., \citealt{heck90}, 
\citealt{shap94}, \citealt{aguirre01}). They are particularly important for 
starburst (SB) galaxies whose intrinsically large star formation rate (SFR) 
provides the perfect environment for gas outflow to develop. 

There has been extensive investigation in the literature on starburst GWs 
(see below).
Our aim here is to extend these studies exploring the  evolution of these winds
in the surrounds of the galaxy. We will address questions which are not yet 
clearly understood such as: (i) how does the feedback of the stars of the 
nuclear region shape the wind? (ii) how much mass is thermally driven? 
(iii) how are the different phases of the gas  connected? (iv) what is the 
fraction of cold  structures built up in the winds? (v) and how does the GW 
impact the chemical evolution of the galaxy? 
In order to try to answer to these questions, we have adopted the galaxy M82 
as a reference to obtain appropriate initial conditions, although the 
results here presented may be applied to others GWs and outflows in general.

M82 is a starburst galaxy, located at about 3.9 Mpc away, with a disk of 
$\sim 5$ kpc radius and mass $\sim 10^{10}$ M$_{\odot}$. 
Although this galaxy has been classified as an Irregular type II, 
several observations \citep{mayya05, barker08, strik09} revealed the presence 
of a bar ($\sim 1$ kpc) and spiral arms (observed in near infrared). 
The main feature of this galaxy is a GW with filamentary structures extending 
up to $\sim 3$ kpc above the disk and an H$\alpha$ cap, i.e. a large 
nebular surrounding feature probably associated with the wind, at a height of 
about 10 kpc \citep{dev_ball99, lehnert99, eng06, strik09}. 
Many studies indicate that this galaxy has experienced a tidal interaction 
with the spiral galaxy M81, resulting a large amount of gas that is being 
piped into M82's core over the last $\sim$ 200 Myr. 
The most recent and important interaction has happened $\sim 2$ - $5 \times
10^{8}$ yr ago, causing an intense starburst which lasted about 50 Myr with 
a SFR $\sim 10$ M$_{\odot}$/yr. Then two more subsequent starbursts occurred, 
and the last one (about 4-6 Myr ago) may have formed at least one of the 
observed largest central star clusters \citep{barker08}.

The SB region in the core of M82 has a diameter of $\sim 1$ kpc and is 
optically defined by portions with high surface brightness labelled A, C, D 
and E by \citet{mangano78}. 
These regions correspond to sources of X-rays, infrared and radio emission. 
The bipolar GW seems to be concentrated within A and C regions and is possibly 
being driven mostly by SNe energy injections.
The SFR leads to a current supernovae rate $\geq$ 0.1 year$^{-1}$. 
The central region has a mass of $\sim 7 \times 10^8$ M$_{\odot}$, 
pressure $\geq$ 10$^{-9}$ dyn cm$^{-2}$, temperature $\sim 10^4$ - 10$^6$ K 
and densities $\sim 10^3$ - 10$^4$ cm$^{-3}$ \citep{cottrell77, mangano78, 
schaaf89, west07, west09a}.
The SNe inject mass and energy at  rates of $\sim$ 1 M$_{\odot}$ yr$^{-1}$ and 
$\sim 2$ $\times 10^{42}$ erg/s, respectively \citep{strik09}.
Recent observations revealed that within a radius of 500 pc there are about 200 
super stellar clusters (SSCs) with an average size of $\sim 5.7$ pc 
and mass (of stars) between 10$^4$ and 10$^6$ M$_{\odot}$ 
\citep{melo05}. 
The projected separation between  each of the young SSCs is smaller than 30 pc,
with a minimum value of 5 pc and a mean separation of $\sim$ 12 pc. 
The density of young SSCs is very high in the M82 starburst galaxy, with a 
global value inferred by \citet{melo05} of 620 kpc$^{-2}$, about 6 times 
larger than that observed in the nuclear regions of others SB galaxies 
(see, e.g., NGC 253, \citealt{watson96}). 
The 20 brightest clusters have ages $\sim 5$ to $10 \times 10^6$ yr 
(\citealt{mccrady03, strik09}).

The bipolar wind of M82 galaxy has a conical geometry with a total opening 
angle of $\sim$ 30 degrees \citep{ranalli08}. 
It has an X-ray luminosity of $\sim 10^{40}$ erg/s, temperatures between 
$\sim 10$ and 10$^8$ K (including the diffuse and warm material, dust and 
molecular gas) and a maximum velocity of the cold filaments (perpendicular to 
the disk)  $\sim 600$ km/s.

The cold and  hot gas metallicities in M82 are uncertain. 
Optical and infrared observations suggest for the cold gas a solar metal 
abundance, while the X-ray observations 
give an abundance for the hot gas $\le$ one-third the solar value 
\citep{strik00}.

To understand the complexity of M82-like GWs many analytical and 
numerical two and three-dimensional studies were performed, most of which
applied to M82 itself \citep{chev85, tenorio98, strik00, tenorio03, 
rod08b, cooper08, strik09}.
\citet{chev85} were the first authors to propose a simple analytical model 
of a steady and adiabatic wind for M82. The same model was later 
revisited  by several authors considering the radiative losses of the ejected 
gas with simplified assumptions. 
\citet{tenorio98} performed two-dimensional simulations of a 
biconical wind with radiative cooling, considering accreted matter from the 
external environment in order to determine the minimum mass of stars and gas 
in the SB that was needed to offset the accreted material. 
\citet{strik00} and, more recently, \citet{strik09}, helped by 
two-dimensional axi-symmetric simulations, studied the dependence of the 
wind dynamics, morphology and X-ray emission  with the distribution of 
the ISM and the star formation history in the SB regions and studied 
also the  distribution of the matter in dense clouds.
Their results indicated that GWs, although efficient at transporting large 
amounts of energy to outside of the galaxies, are inefficient at transporting 
mass. \citet{tenorio03} investigated also by means of two dimensional 
simulations the formation of filaments in a wind driven by SNe from  
several star clusters, 
while \citet{cooper08} used three-dimensional (3D) simulations to study for
the first time the formation of the wind in an inhomogeneous disk.
They found that the presence of an inhomogeneous ISM in the disc may have an 
important effect on the morphology of the wind itself and on the distribution 
of the filaments.
Finally, \citet{rod08a,rod08b} also by means of 3D simulations
have modeled the filaments in a GW as the result of the interaction between 
the winds from a distribution of SSCs and derived the condition necessary for 
producing a radiative interaction between the cluster winds, as in M82.
In general, all these numerical studies indicate a coexistence in the wind 
of a low density, high temperature gas which expands freely 
and occupies a large volume with a cooler gas (with a filamentary 
structure which emits H$\alpha$ lines) that expands more slowly
and occupies a small fraction of the total volume.
These models also predict a velocity for the hot phase 
(T $\le$ 10$^8$ K and n $\ge$ 10$^{-4}$ cm$^{-3}$) of $\sim 1000$ - 2240 km/s, 
and for the cold phase (T $\leqslant$ 10$^4$ K and n $<$ 10$^2$ cm$^{-3}$) of 
$\sim 600$ km/s, in agreement with the observations.

In this paper these previous analyses will be extended through a detailed 
numerical study of a SN-driven M82-like wind in the surrounds of 
the ejection region of the galaxy.
We  perform  fully three-dimensional (3D) hydrodynamical simulations with 
radiative cooling in a gas in
collisional ionisation equilibrium (CIE). The cooling function is 
computed taking into account  the metal abundances 
both in the ISM and in the SNe ejecta. As in the previous 
studies of \citet{tenorio98}, \citet{tenorio03} and \citet{rod08b}, 
in our model the wind is generated by the ejected energy of the SNe hosted 
in the SSCs which are randomly distributed in time and space 
within a nuclear region of radius 150 pc. However, in contrast to 
those studies, we employ a 3D galactic disk geometry with a multi-phase 
stratified and rotating ISM which is particularly important to assess 
realistically the multi-phase complex wind's non-axisymmetric structures.
Furthermore, with a more realistic treatment of the gas radiative cooling 
we will be able to track the formation, evolution and stability of the dense 
and cold structures (filaments), which are formed as a result of the 
interactions between the dense radiatively cooled shells of the several 
supernova remnants (SNRs) generated in the SB, and coexisting in equilibrium 
with the high temperature, low density wind component, as well as to determine 
more appropriately the fraction of the whole volume of the wind which is 
occupied by this cold phase. 
Also, the evolution of the SN ejected matter, (which is traced by a 
separated continuity equation as a passive scalar) will allow us to obtain 
for the first time in numerical simulations of GWs important information 
on the chemical evolution, as well as of the H$\alpha$ and the soft X-ray 
emission of the wind, and the metal contamination of the intergalactic 
medium (IGM) due to the wind itself.

In the following sections we will outline the main characteristics of our model
(Sec. 2) and the results obtained from the numerical simulations, mostly 
highlighting the formation process of the filaments, the chemical evolution of 
the wind and the distribution of the most abundant species outside of the disk 
(Sec. 3). Finally, in Sec. 4 we will present a brief discussion and 
draw our  conclusions, and in Sec. 5 a summary of the main results found in 
this work.

\section{The model}

\subsection{Setup for an M82-like starburst wind}

In our model, the gas of the disk is initially set in rotational equilibrium 
in the gravitational potential of the galaxy. The gravitational potential of 
the stars, $\Phi_{star}(r,z)$ is assumed to be generated by a stellar 
distribution with  a spheroidal King profile
\begin{equation}
\label{eq:rhodisk}
\rho_{\star}(r,z)={\rho_{\star,0} \over \left [1+ (r^2+z^2)/\omega_0^2 
\right]^{3/2}}.
\end{equation}
\noindent
where $\rho_{\star,0}$ is the central density of the stars, and
$\omega_0=\sqrt (r_0^2+z_0^2)$=350 pc is the core radius. 
With this  distribution the stellar gravitational potential is:
\begin{equation}
\label{eq:rhodisk}
\Phi_{star}(r,z) = - \frac{GM_{star}}{\omega_0} \left[\frac{ln 
\left \{(\sqrt{r^2+z^2}/\omega_0) + \sqrt{1+(r^2+z^2)/\omega_0^2} \right \}}
{\sqrt{r^2+z^2}/\omega_0} \right]
\end{equation}
\noindent
where $M_{star}$=$4 \pi \rho_{star} \omega_0^3$= 2 $\times 10^8$ 
M$_{\odot}$ is the stellar mass within the radius $\omega_0$.

The gravitational potential of the gas disk is given by Miyamoto \& Nagai 
model (1975), that is:

\begin{equation}
\label{eq:rhodisk}
\Phi_{disk}(r,z) = - \frac{GM_{disk}}{\sqrt{r^2 + \left(a+
\sqrt{z^2+b^2}\right)^2}}
\end{equation}
\noindent
where $a=222$ pc and $b=75$ pc are the radial and vertical scales of the disk,
respectively, and $M_{disk}=2 \times 10^9$ M$_{\odot}$. In our model we have
$\Phi(r,z)$ = $\Phi_{star}(r,z)$ + $\Phi_{disk}(r,z)$. We do not 
incorporate a dark matter halo component because we are interested in the 
behaviour of the wind evolving for only a short period of time 
($\sim$ 10 Myr) and within a small volume in the surroundings of the galaxy, 
so that the gravitational effects of a dark matter 
halo are negligible in this case \citep{strik00}.

For an initially isothermal gas, the disk  density distribution will therefore 
have the form:

\begin{equation}
\label{eq:disk}
\rho_{d}(r,z) = \rho_{d,0} \; exp \left[-\frac{\Phi(r,z) - 
e^2\Phi(r,0)-(1-e^2)\Phi(0,0)}{c_{s,T,disk}^2}\right]
\end{equation}
\noindent
where $c_{s,T,disk}^2$ is the isothermal sound speed in the disk and 
$e=e_{rot}^{(-z/z_{rot})}$ quantifies the fraction of rotational support of the 
ISM. To reduce the rotational velocity of the gas above the plane, we have 
assumed a simple model where the rotational support drops off exponentially 
with increasing height $z$ and where the scale height for this reduction in the
rotational velocity is $z_{rot}$ = 5 kpc (see \citealt{strik00}). 
Similarly, the distribution of the gas of the halo is assumed to be of the
form:

\begin{equation}
\label{eq:disk}
\rho_{h}(r,z) = \rho_{h,0} \; exp \left[-\frac{\Phi(r,z) - 
e^2\Phi(r,0)-(1-e^2)\Phi(0,0)}{c_{s,T,halo}^2}\right]
\end{equation}
\noindent
where $c_{s,T,halo}^2$ is the isothermal sound speed in the halo.
We adopted a disk temperature, $T_{disk} \sim 
6.7 \times 10^4$ K and a halo temperature $T_{halo} \sim 6.7 \times 10^6$ K.
Therefore, in each point of the system the total density is $\rho(r,z) = 
\rho_d(r,z)+\rho_h(r,z)$ and the total pressure is $p(r,z)=\rho_d(r,z) 
c_{s,T,disk}^2 + \rho_h(r,z) c_{s,T,halo}^2$. The galaxy parameters 
adopted in this study are presented in Table \ref{tab:gal}.

\begin{table*}
 \centering
 \begin{minipage}{140mm}
  \caption{Parameters for the galaxy setup obtained from M82}
  \label{tab:gal}
  \begin{tabular}{@{}ccccccccccc@{}}
  \hline
 $M_{star}$ & $M_{disk}$ & $\omega_0$ & $a$ & 
$b$ & $c_{s,T,disk}$ & $c_{s,T,halo}$ 
& $e_{rot}$ & $z_{rot}$ & $\rho_{disk,0}$ 
& $\rho_{halo,0}$ \\
M$_{\odot}$ & M$_{\odot}$ & pc & pc & pc & cm/s & cm/s & & kpc & cm$^{-3}$ 
& cm$^{-3}$ \\
 \hline
 2$\times 10^8$ & 2$\times 10^9$ & 350 & 222 & 75 & 30$\times 10^5$ & 
300$\times 10^5$ & 0.95 & 5 & 20 & 2$\times 10^{-3}$ \\
\hline
\end{tabular}
\end{minipage}
\end{table*}

\subsection{Energy injection}

The SB activity in M82 is centred in the nuclear region with a 
diameter of $\sim$ 500 pc. As described above, within this region there is 
a number of prominent and high surface brightness clumps containing  
hundreds of young massive SSCs \citep{connell95, melo05}.
In our study we assume that the kpc-scale bipolar superwind observed in 
different wavelengths \citep{shop98, steve03, strik07} 
is driven by the SN explosions occurring in the SSCs, and that the total 
luminosity injected into the ISM is $\sum_1^{\cal N_{SSC}} L_i= 10^{42}$ erg 
s$^{-1}$, where $L_i$ is the luminosity of a single SSC and ${\cal N}_{ssc}$ is 
the total number of SSCs.
We also  assume that ${\cal N}_{ssc}$ ranges between a minimum of 10 
and a maximum of 100 and consider an inner nuclear region within a radius 
($R_{SB}$) of 150 pc where this activity takes place. 
We further consider that the total amount of SSCs may be 
formed in a star formation process along 1 or 10 Myr, depending on the model.
Therefore, for each cluster we assume an average luminosity  
$L_i = 10^{42}/{\cal N}_{ssc}$ erg s$^{-1}$ and, since every SN injects 
$E_0=10^{51}$ erg, we have ${\cal N}_{SN,c}$ = 
$10^{42}$(t$_b/E_0 {\cal N}_{ssc}$) = $10^6$/${\cal N}_{ssc}$,
where ${\cal N}_{SN,c}$ is the total number of SNe per cluster, and where we 
assume that each single cluster is formed in an instantaneous burst which is 
characterized by SN activity occurring over a time t$_b$ = 30 Myr 
(i.e. the lifetime of the less massive SN II progenitor star) and by a 
SN heating efficiency HE = 100\% \citep{mel04}, where the heating efficiency, 
HE, corresponds to the fraction of the SN explosion energy that remains 
effectively stored in the ISM gas (as kinetic and thermal energies) and is not 
radiated away.\footnote{We note that due to the short lifetime of the 
starburst, almost all the injected energy in the wind is due to SNe II only. 
For this reason, in this study we neglect the contributions due to type I SNe 
explosions.} 
Therefore, an SSC injects mass and energy at rates $\dot M$ = 
M$_{\rm ej}$ $10^{42}$/($E_0 {\cal N}_{ssc}$) and $L_{\rm w}$ = 
$10^{42}/{\cal N}_{ssc}$, respectively, where M$_{\rm ej}$=
16 M$_{\odot}$ is the mean mass released by a single SN explosion.
According to \citet{marc07}, we also assume that each SN explosion injects 
3 M$_{\odot}$ of metals into the ISM.
Finally, to set the clusters spatially and temporally within the core
of the galaxy, we associate randomly to each $i$-th cluster a position $P^i=
(r_i,\phi_i)$, where $0<r_i<150$ pc and $0<\phi_i<2 \pi$, and a time 
$t^i$ in the range $0<t^i<1$ Myr or $0<t^i<10$ Myr depending on the model. 

\begin{figure}    
\begin{center}   
\psfig{figure=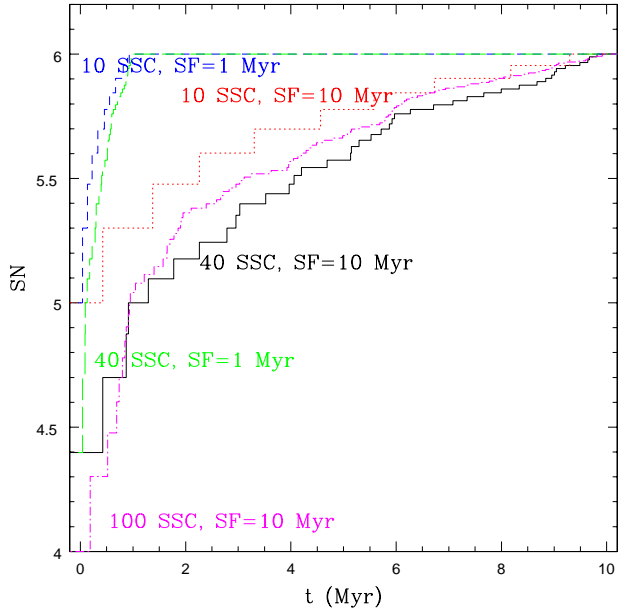,width=0.49\textwidth}    
\end{center}   
\caption{Number of SN explosions in the first 10 Myr for each model. 
Solid (black) line: SC40-10; dashed (blue) line: SC10-01; long-dashed (green) 
line: SC40-01; dotted (red) line: SC10-10 and dotted-dashed (magenta) line: 
SC100-10}
\label{fig:sn_inject} 
\end{figure}

A summary of all the parameters adopted for the intermittent wind models 
studied here are given in Table \ref{tab:mod} and Figure \ref{fig:sn_inject} 
shows histograms representing the number of SN explosions during the first 
10 Myr for each  model.

\subsection{Numerical methodology}

To simulate the evolution of the superwind driven by SN explosions in an
M82-like SB galaxy we use a modified version of the numerical adaptive 
mesh refinement  hydrodynamical code YGUAZU \citep{raga00, raga02, mel08I, 
mel09II} that integrates the gas-dynamic equations with the $flux$ 
$vector$ $splitting$ algorithm of Van Leer (1982). 
It also includes a parametrized cooling function in the energy equation 
that allows the gas to cool from $\sim$ 10$^{7.5}$ to 10$^4$ K with errors 
smaller than 10$\%$ and which is calculated implicitly in each time step for 
each grid position.
The 3D binary, hierarchical computational grid is structured with
a base grid and with a number of nested grids whose resolution
doubles going from one level to the next one. 
We have run the models with a maximum resolution of 5.8 pc per cell.
\footnote{We have also performed several tests with a larger resolution of 2.9 
pc per sell for 2 Myr and found that the results were very similar to those 
with a 5.8 pc resolution per cell. For this reason, we adopted the latter 
scheme for the study presented here as it allowed us to save computation time 
without loosing information and  quality in the results.}
For one model (SC40-01), which we take as the reference model, we have 
adopted five grid levels covering a box with physical dimensions of 
1.5$\times$1.5$\times$3.0 kpc in the x, y and z directions, respectively, 
while for all the remaining  models we have considered a box with physical 
dimensions of 1.5$\times$1.5$\times$1.5 kpc. 
The maximum dimensions of the computational domain for each model are also 
presented in Table 2.
The energy associated to each SSC is injected as thermal energy in a 
single cell of the highest grid level of the box. In order to account for 
the metal abundance injected by the SN explosions into the system, we have 
introduced an extra separated continuity equation into the code 
in the form ${\rm d}\rho_z /{\rm d}t + \nabla 
\cdot(\rho_z v)=S_z$, where $\rho_z$ is the metal density, and $S_z$ is the 
SN metal source.
This also allows us to easily track the metallicity evolution in 
the wind.

Finally, for computing the radiative cooling we employed a cooling function 
$\Lambda$(T) considering an optically thin gas in the ionization equilibrium 
regime, which implies 
\begin{equation}
{\cal L}(\rho, T) = \rho^2 \Lambda(T) Z \ \ {\rm erg} \ {\rm cm}^{-3} 
{\rm s}^{-1} 
\end{equation}
\citep{mcwhirt75, aldro73, degouv93}, where $Z$ is the
total gas abundance computed at each time and at each point of the grid.

\begin{table*}
 \centering
 \begin{minipage}{140mm}
  \caption{Wind parameters used in our hydrodynamic simulations for
5 different models. For each Model: $a)$ Number od SSCs; $b)$ 
Duration of the SF process (in Myr); $c)$ Injected luminosity $per$ SSC 
(in erg/s); $d)$ Injected matter $per$ SSC (in M$_{\odot}$/yr); 
$e)$ Total energy injected after the first 5 Myr (in units of $10^{55}$ erg); 
and $f)$ Physical dimensions of the box (in kpc).}
  \label{tab:mod}
  \begin{tabular}{@{}llrrrrr@{}}
\hline
  Model & SSCs ($\sharp$)    & SF (Myr)   & $\dot{E}$/SSC (erg/s) & 
  $\dot{M}$/SSC (M$_{\odot}$/yr) & $E_{5}$ ($10^{55}$erg) & Box Size (kpc) \\
\hline
  SC10-10  & 10  &   10 &    $10^{41}$          &  $5\cdot 10^{-2}$ & 3.2   
& 1.5 $\times$ 1.5 $\times$ 1.5 \\
  SC10-01  & 10  &    1 &    $10^{41}$          &  $5\cdot 10^{-2}$ & 7.3   
& 1.5 $\times$ 1.5 $\times$ 1.5\\
  SC40-10  & 40  &   10 &    $2.5\cdot 10^{40}$ &  $1.25\cdot 10^{-2}$ & 1.6  
& 1.5 $\times$ 1.5 $\times$ 1.5\\
  SC40-01  & 40  &    1 &    $2.5\cdot 10^{40}$ &  $1.25\cdot 10^{-2}$ & 7.2 
& 1.5 $\times$ 1.5 $\times$ 3.0\\
  SC100-10 & 100 &   10 &    $10^{40}$          &  $5\cdot 10^{-3}$ & 2.1  
& 1.5 $\times$ 1.5 $\times$ 1.5\\
\hline
\end{tabular}
\end{minipage}
\end{table*}

\section{Results}

In this section we present the results for the different models listed 
in Table \ref{tab:mod}, exploring a set of SSC parameters aiming mostly at 
the examination of the injected energy, the multiphase ISM and the chemical 
evolution of the system. 
All the  models, except the reference one (SC40-01), were evolved for 5 Myr
right after the first stars in an SSC (with masses 120 M$_{\odot}$) 
have begun to explode,
which is an appropriate time to investigate the global properties of the wind 
build up, from the first SN explosions up to a nearly steady state phase 
reached after $\sim$ 3.5 Myr. The reference model, instead, 
was evolved for 8.5 Myr, in order to follow the gas outflow evolution up to a 
height of 3 kpc.
The results here obtained may be directly compared with observations of 
GWs in order  to determine, for instance, the best set of 
initial conditions able to drive their outflows.

Table \ref{tab:ene} summarizes the global energy budget for all the models 
after the reference times  t=3.5 and t=5 Myr. 
These energies will help us to understand the wind evolution in the 
discussion below.

\begin{table*}
 \centering
 \begin{minipage}{140mm}
  \caption{Energy budget  (in units of $10^{55}$ erg) at t=3.5 Myr
($E_{3.5}$) and t=5 Myr ($E_{5}$). For each Model: $a)$ energy injected by the 
SN explosions; $b)$ total energy stored into the system; $c)$ total kinetic
energy; $d)$ vertical (or poloidal) kinetic energy; $e)$ total thermal energy; 
$f)$ total energy lost by radiative cooling; $g)$ fraction of the injected SN 
energy which is radiated away.}
  \label{tab:ene}
  \begin{tabular}{@{}lrrrrrrr@{}}
\hline
  Model & E$_{SN,3.5}$ & E$_{sy,3.5}$ & E$_{k_{tot},3.5}$ & E$_{k_{z},3.5}$ & E$_{th,3.5}$
  & E$_{lost,3.5}$ & E$_{rad,3.5}$ (\%) \\
\hline
  SC10-10  & 2.20 & 1.10 & 1.28 & 0.45 & 0.20 & 1.50 & 68 \\
  SC10-01  & 5.60 & 2.35 & 2.00 & 0.98 & 0.33 & 4.10 & 73 \\
  SC40-10  & 0.86 & 1.18 & 0.98 & 0.20 & 0.19 & 0.53 & 62 \\
  SC40-01  & 4.90 & 2.80 & 2.46 & 1.50 & 0.39 & 2.95 & 60 \\
  SC100-10 & 1.10 & 1.14 & 0.95 & 0.10 & 0.23 & 0.81 & 74 \\
\hline
\hline
  Model & E$_{SN,5}$ & E$_{sy,5}$ & E$_{k_{tot},5}$ & E$_{k_{z},5}$ & E$_{th,5}$
  & E$_{lost,5}$ & E$_{rad,5}$ (\%) \\
\hline
  SC10-10  & 3.20 & 1.75 & 1.49 & 0.42 & 0.26 & 2.30 & 72 \\
  SC10-01  & 7.30 & 2.15 & 1.90 & 1.09 & 0.30 & 6.00 & 82 \\
  SC40-10  & 1.60 & 1.25 & 1.08 & 0.26 & 0.17 & 1.20 & 75 \\
  SC40-01  & 7.20 & 2.70 & 2.35 & 1.30 & 0.35 & 5.35 & 74 \\
  SC100-10 & 2.10 & 1.30 & 1.04 & 0.28 & 0.26 & 1.65 & 79 \\
\hline
\end{tabular}
\end{minipage}
\end{table*}

\subsection{Model SC10-10}

The first model (SC10-10) considers 10 SSCs formed along 10
Myr, each of them having  a total stellar mass of $10^7$ M$_{\odot}$, 
a total SNe number of $10^5$ and a mean separation of $\sim$ 50 pc. 
With such configuration, the SN explosions occurring in the first 
SSCs have more than sufficient energy to drive an outward propagating shock 
wave which sweeps up the ISM very quickly. In fact, as a single SN 
explosion generates a supernova remnant (SNR), the whole set of  SNe 
explosions  due to the first SSCs generates several SNRs which interact with 
each other still in the nuclear region of the disk and drive an outward 
propagating large scale shock front. 
This sweeps up the ISM into a thin and dense shell that envelopes a large, 
hot, low-density cavity.
Such structure, usually denominated superbubble, has been also 
detected in former numerical studies of GW formation (see, e.g., 
\citealt{tomi81, tomi86}). 
It is maintained at a high temperature and a low density due to the continuous 
injection of SN energy. 
New  superbubbles generated by later SSCs will expand into 
such a rarefied ISM without forming new dense shells and thus 
suffering negligible radiative losses. 
Therefore, with the setup of model SC10-10, almost all the SN energy 
is transferred to the ISM without being radiated away and the result 
is a highly energized superbubble driven by the 
energy ejected from each SSC, without forming a wind with filamentary 
structures immersed in it. 
Nonetheless, at least the shock compressed external shell that 
develops in the beginning is radiativelly cooled and suffers fragmentation 
due to the combination of the development of the Kelvin-Helmholtz (K-H) and 
Rayleigh-Taylor (R-T) instabilities as the dense gas of the shell is 
pushed into the rarefied gas of the disk halo
(e.g. \citealt{meli05, deGBenz93}).

\begin{figure}
\begin{center}   
\psfig{figure=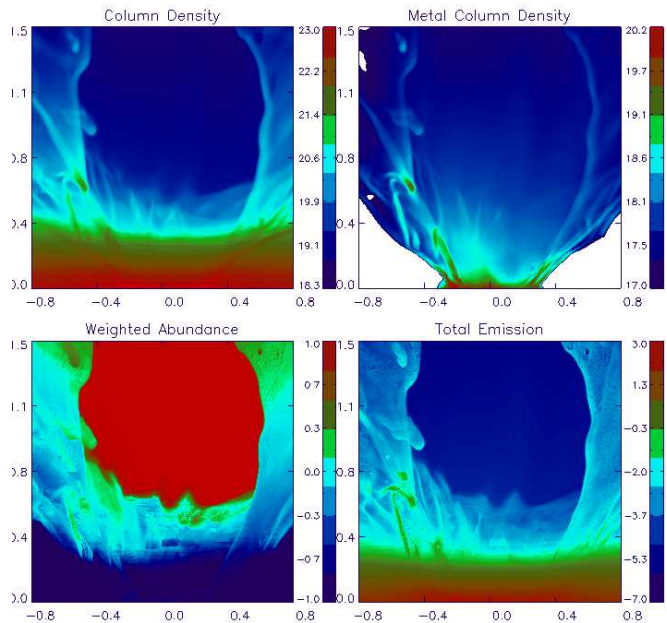,width=0.49\textwidth}    
\end{center}   
\caption{Model SC10-10: color-scale map of the gas column density (top-left 
panel), the metal column density (top-right), the total weighted abundance, 
(i.e. $\sum_{y=0}^{y_{max}}{\rho_{Z}(x,z)}/\sum_{y=0}^{y_{max}}{\rho(x,z)}$, where 
$\rho_{Z}(x,z)$ and $\rho(x,z)$ are the metal and total gas
density in the space coordinates $x,y$, respectively)( bottom-left) 
and the total emission (bottom-right) at at a time t=3.8 Myr.
The column density is shown in units of cm$^{-2}$, the abundance is normalized 
by the solar abundance and the emissivity is in erg s$^{-1}$. 
All the quantities are expressed in log-scale. In this model, the SSCs have a 
mass of $10^7$ M$_{\odot}$, a SN number of $10^5$ and are formed in a starburst 
process of 10 Myr.}
\label{fig:m82_10-10} 
\end{figure}

This is the resulting scenario that we see in Figure 
\ref{fig:m82_10-10} after 3.8 Myr when the upper part of the wind has 
already left the computational domain.
The top-left panel shows the gas column density  and we note 
that the only dense structures are formed along the expanding shocked shell.
No filaments or clouds are formed within the wind flow and consequently
the gas of the disk carried out by the wind and lost to the IGM is essentially 
the interstellar gas swept by the first generation of SN explosions. 
This is also suggested by the top-right and bottom-left diagrams of Figure 
\ref{fig:m82_10-10}.
The low density phase of the wind, which is enriched by the metals ejected 
by the SNe, has a high abundance $\sim$ 10 times larger than the solar 
abundance.
Nonetheless, most of the metals remain in the nuclear region of the galaxy. 
In this case we also note that the thermal energy radiated by the wind
comes from the external shell, characterized by a temperature of $\sim$ 
10$^4$ K (bottom-right panel of Figure \ref{fig:m82_10-10}). 

Therefore, the results above suggest that when the gas outflow is 
generated by only a few very massive SSCs most of the soft and/or hard X-ray 
emission will be detected only behind the shocked regions of the shell-like 
structure around the GW, provided that the reverse shock where the 
wind impacts the outward propagating shock is right behind the later.

\begin{figure}    
\begin{center}   
\psfig{figure=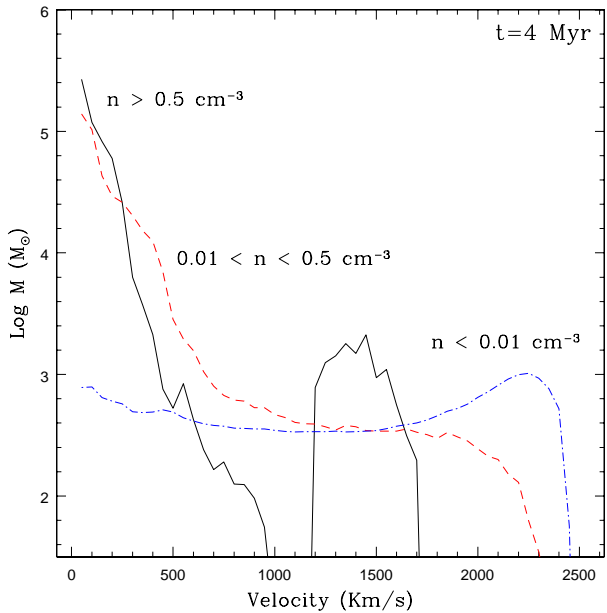,width=0.49\textwidth}    
\end{center}   
\caption{Model SC10-10: Gas mass distribution as function of the vertical 
velocity for three different phases of the gas at t=3.8 Myr. 
Solid line: gas with number density larger
than 1 cm$^{-3}$; dashed line: gas with number density between $10^{-2}$ and 1 
cm$^{-3}$; dotted-dashed line: gas with number density smaller than $10^{-2}$ 
cm$^{-3}$. The velocity is expressed in km s$^{-1}$ and the mass is in units of 
solar mass (in log-scale). In this model the SSCs were formed over a period of 
10 Myr.}
\label{fig:V_10-10} 
\end{figure}

\begin{figure}
\begin{center}
\psfig{figure=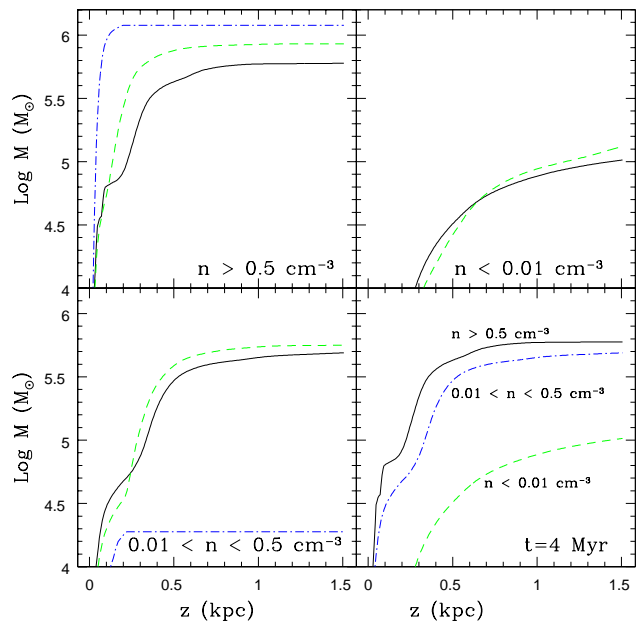,width=0.49\textwidth}
\end{center}
\caption{Model SC10-10: Total gas mass distribution as function of $z$ (height above the disk)
at three different times: t=1.5 Myr (dashed line), t=2.5 Myr 
(dotted-dashed line) and t=3.8 Myr (solid line). 
Top-left panel: gas with number density 
larger than 1 cm$^{-3}$; top-right panel: gas with number density between 
$10^{-2}$ and 1 cm$^{-3}$; bottom-left panel: gas with number density smaller 
than $10^{-2}$ cm$^{-3}$. 
Bottom-right panel: comparison between the total gas mass 
at t=3.8 Myr in the high (solid line), intermediate (dashed line) and low 
(dotted-dashed line) density phases of the gas. Mass is expressed in M$_{\odot}$ 
(log-scale) and $z$ is in kpc.}
\label{fig:fil_10-10}
\end{figure}

Investigating carefully the features of the gas, we
realize that in  model SC10-10 there are two well separated gas phases. 
The first one associated to the internal part of the superbubble which has 
 a density smaller than 10$^{-1}$ cm$^{-3}$,
a temperature of $\sim$ 10$^7$ K and a velocity between 100 and 2000 km 
s$^{-1}$. The other gas phase is associated to the external shell of
the superbubble, with a density larger than 10$^{-1}$ cm$^{-3}$, a temperature of
$\sim$ 10$^4$ K and a velocity normal to the disk direction 
of $\sim$ 1500 km s$^{-1}$.
These features can be also distinguished in Figure \ref{fig:V_10-10}, where we
have plotted the gas mass as function of its vertical velocity,
for three different ranges of gas densities, and in Figure \ref{fig:fil_10-10},
where the total mass of each phase with a vertical velocity $\ge$ 50 km 
s$^{-1}$ is plotted as function of $z$ (the height above the disk).
We observe that the intermediate-phase, characterized by a density of 
$\sim$ 10$^{-1}$-10$^{-2}$ cm$^{-3}$, has a similar mass to the high-density 
phase, of $\sim$ 3$\times 10^5$ M$_{\odot}$. Its velocity follows
the velocity distribution of the high-density phase
up to $\sim$ 500 km s$^{-1}$ and then, follows the distribution of the
low density phase. This means that its evolution does not occur separately
from the other phases. In the disk (where the velocities are
smaller), this phase is related to the high density phase, while in the wind
(where the velocities are larger), it is connected to the low density phase.
Figure \ref{fig:V_10-10} also highlights the presence of a new generation of 
accelerated shells which is represented by the peak  velocities between 
1200 and 1700 km s$^{-1}$. It is well separated from the high density gas, 
which has velocities between 50 and 900 km s$^{-1}$ and is probably distributed
along the external shell pushed by the outflow.
The total mass of this new generation of shells is small, because almost all 
the gas of the nuclear region has been already swept by the first generation 
of SN explosions and therefore, no filaments or structures are driven along 
the wind.

The results above indicate that the SN-driven gas outflow from the nuclear 
region of a galaxy with only a few highly massive SSCs radiates about
68\% of the total energy injected by the SNe.
In fact, after 3.5 Myr, when the wind has already  filled completely the 
physical domain of our simulation, 
the sum of the kinetic and thermal energies of the system is $\sim$ 1.5 
$\times 10^{55}$ erg, i.e. almost twice the energy of the system at $t=0$, 
corresponding to 8.5 $\times 10^{54}$ erg, while the total energy injected by 
the SN explosions is about 2.2 $\times 10^{55}$ erg. 
Therefore, nearly 2/3 of the SN energy is radiated away 
(mainly in the nuclear region and in the external shell), and the remaining 
1/3 goes to the kinetic energy (20\% of the total injected energy) and thermal 
energy (10\% of the total injected energy) of the gas. 
This result is consistent with the evolution of the multi-phase ISM shown in 
Figure \ref{fig:fil_10-10}. Over the simulated period, the amount of
gas with a cooling time shorter than the dynamical time, represents $\sim$ 
60\% of the total gas flow and thus assuming that the SN energy is spread 
approximately uniformly, it is reasonable to believe that all the energy in 
the shells is radiated away, while the remaining energy in the low and 
intermediate-density gas phases increases the gas pressure and consequently, 
the thermal and kinetic energy of the GW.

\subsection{Model SC10-01}

\begin{figure}
\begin{center}
\psfig{figure=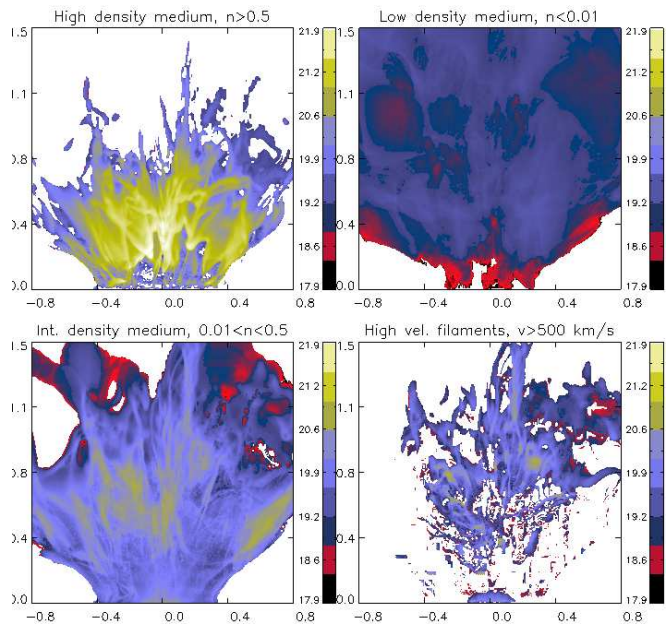,width=0.49\textwidth}
\end{center}
\caption{Model SC10-01: map of the gas column density for different density
and velocity ranges at a time of 3 Myr.
top-left panel: $n \ge 0.5$ cm$^{-3}$ and $v \ge 50$ km s$^{-1}$; top-right
panel: $0.01 \le n \le 0.5$ cm$^{-3}$ and $v \ge 50$ km s$^{-1}$; bottom-the
left panel: $n \le 0.01$ cm$^{-3}$ and $v \ge 50$ km s$^{-1}$; bottom-right
panel: $n \ge 0.5$ cm$^{-3}$ and $v \ge 300$ km s$^{-1}$. The column density is
expressed in cm$^{-2}$. The "white" portions in each panel represent regions
with total column density smaller than 8$\times 10^{17}$ cm$^{-2}$.}
\label{fig:m82_10-1wp}
\end{figure}

The second Model (SC10-01) considers, as in the previous one, 10 SSCs with a
stellar mass of $10^7$ M$_{\odot}$,
a SNe number of $10^5$ and a mean separation of $\sim$ 50 pc. However, in this
case we assume that all the super clusters are formed along  1 Myr only and
therefore, in contrast with the first model, every SSC develops a hot bubble
surrounded by a thin dense shell at about the same epoch.
The SN explosions occurring in the SSCs have
sufficient energy to drive the bubble expansion at the same time that the
collisions between the shells lead to the formation of an interconnected
network of dense filaments permeated by tunnels containing very hot, tenuous
ISM. In fact the SNRs generated by the SN explosions propagate into an 
ambient medium with a density $\ge$ 1 cm$^{-3}$, and enter in the radiative 
phase in a time shorter than 3$\times 10^4$ yr \citep{mc87}, that is, at 
radii smaller than $\sim$20 pc \citep{mel06}. 
Therefore, their external shells are expected to become very dense 
($n_{sh}/n \ge 10$) and cold (T $\le 10^4$ K) before interacting with each 
other. Collisions among them lead to the formation of very dense filaments 
(up to 1000 times denser than the hot gas) and dense surfaces (up to 100 
times denser than the hot gas) in times between $10^5$ and $10^6$ yr according 
to our simulation. R-T and K-H instabilities are 
also able to fragment these structures and produce new filaments, in a process 
that can continue during all the SN activity, until the average gas density of 
the SB is $\geq$ $10^{-1}$ cm$^{-3}$ (see e.g., \citealt{mc87} and 
\citealt{mel06}).
In spite of the initial formation of filaments and dense structures, the 
superbubble resulting from the interaction
of the individual SSC bubbles evolves as in Model SC10-10 and when also the
external shell fragments due to R-T and K-H instabilities, the hot gas is
able to escape from the galaxy without carrying very large amounts of
metals or disk gas.
Nonetheless, Figure \ref{fig:m82_10-1wp}, which maps the column density
edge-on distributions of each gas phase in different velocity ranges,
clearly shows that a bunch of filaments and
a large amount of metals launched together with the hot and low density
gas (top-left and top-right panels of Figure \ref{fig:m82_10-1wp})
and consequently, a large fraction of energy is radiated away by the
filaments themselves.

\subsection{Model SC40-10}

Now we are going to analyse the results of our simulations considering a SB 
region with intermittent injection of mass and energy from a larger number of 
SSCs (40) having a stellar mass of $2.5 \times 10^6$ M$_{\odot}$, a SNe number 
of $2.5 \times 10^4$ and a mean separation of $\sim$ 40 pc. 
In this case we assume, as in Model SC10-10, that the SSCs are formed
over a time of 10 Myr (Model SC40-10).  

The resulting  gas outflow is very similar to the one obtained in  
Model SC10-10
but it has an amount of filaments up to a height $\sim$ 600 pc. 
Also in this case, despite the high metal abundance of the wind flow 
($\sim$ 10 Z$_{\odot}$), most of the metals ejected by the SNe remain in the 
disk, and only a small fraction is transported to outside of 
the system together with the large external shell formed by the activity of 
the first SSCs.
Finally, the gas emission of the central region (up to 500 pc) is about
1000 times larger than the emission of the wind and 100 times larger than
the emission of the external shell. 
Therefore, in such a case, we expect an X-ray emission
coming only from the SB core of the galaxy and a weak radio and infra-red 
emission from the shell fragments at $\sim$ 1.5 kpc from the disk, 
characterized by a maximum temperature of $\sim$ 10$^4$ K.

Although the main features of the wind are different from those observed in
M82, we note that an increase in the number of SSCs from model SC10-10 to 
model SC40-10 favours the formation of a multi-phase gas in the latter.
The dense, cold gas phase shows a 
peak between velocities of 600 and 800 km s$^{-1}$, corresponding to a total 
gas mass of about 3000 M$_{\odot}$ in accelerated filaments in the nuclear 
region.
At the same time, both the intermediate and low density gas are accelerated up 
to velocities of 2000 and 2300 km s$^{-1}$, respectively, and the intermediate 
(warm)-phase velocity distribution overlaps with that of the denser 
(cold)-phase until 800 km s$^{-1}$, which corresponds to the maximum velocity 
of the filaments.

To understand better the multi-phase gas distribution, as before, we have 
analyzed the edge-on column density of each gas 
phase, which shows that in the first 500 pc the cold (high density),
warm (intermediate density), and hot (low density) gas phases are completely
mixed up. Above 500 pc, instead, the phases are
decoupled, and no clouds or filaments are observed together with the low
density launched flow.

In this model the energy radiated away is about 75\% of the total energy 
injected by the SNs. After 5 Myr the sum of the kinetic and thermal energies 
of the system is $\sim$ 1.25 $\times 10^{55}$ erg, while the total injected 
energy is $\sim$ 1.6 $\times 10^{55}$ erg. This means that about 3/4 of the SN 
energy is lost by radiative cooling, while the remaining 1/4 goes into the 
kinetic energy (15\% of the total injected energy) and thermal energy (10\% 
of the total injected energy) of the gas. 
Also in this case, the energy distribution is consistent with the composition 
of the multi-phase ISM. 
After 4 Myr, the amount of gas in the intermediate and high density phase 
represents $\sim$ 65\% of the total gas flow, but we note that earlier this 
fraction was larger than 80\%.
This means that the energy was radiated away with high
efficiency and proportionally to the total amount of gas with density larger
than 0.01 cm$^{-3}$.

\subsection{SC40-01 - the reference Model}

The fourth Model (SC40-01) considers 40 SSCs (as in Model SC40-10) with a 
stellar mass of $2.5 \times 10^6$ M$_{\odot}$, a SNe number of $2.5 \times 10^4$,
a mean separation of $\sim$ 40 pc, but with the formation of the SCCs over 1 
Myr only.
This model, which we denominate our Reference Model, due to the larger energy 
power injected (see Table 3) is able to produce and launch a multi-phase wind 
to greater distances. For this reason, we have run this simulation in a larger 
box domain with physical dimensions of 1.5 $\times$ 1.5 $\times$ 3 kpc.

\begin{figure}    
\begin{center}   
\psfig{figure=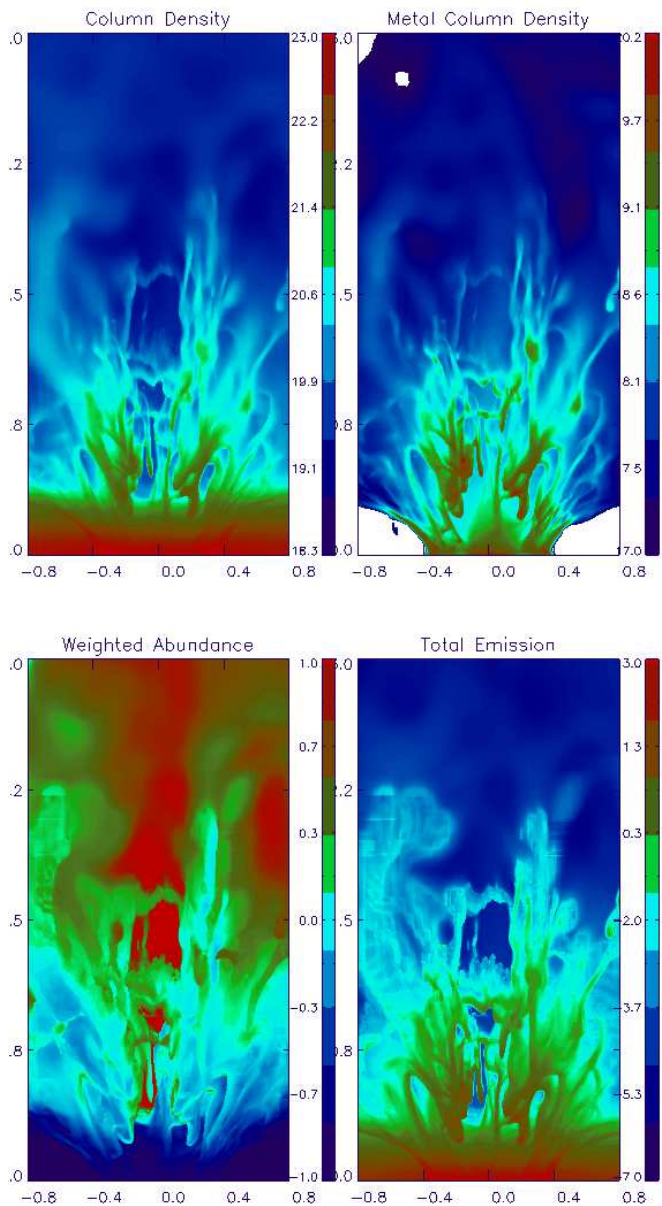,width=0.49\textwidth}    
\end{center}   
\caption{Model SC40-01: the same as in Fig. \ref{fig:m82_10-10}, at t=3.5 Myr,
except that here we consider 40 SSCs forming over 1 Myr with a mass of 2.5 
$\times 10^6$ M$_{\odot}$ and a SN number of 2.5 $\times 10^4$.}
\label{fig:m82_40-1} 
\end{figure}

As described in Model SC-10-01, the larger number of SSCs and the 
shorter time (1 Myr) at which they begin to be active (which lead to a larger 
injected energy power), have created the conditions to produce an environment 
with rich filamentary structure and with a thermal pressure large enough to 
drive both the low and high density phases of the wind very quickly to large 
distances. In a few Myr, a multi-phase ISM is launched in the IGM and a GW 
with a large amount of dense structures embedded in a hot low density gas 
develops.
Figure \ref{fig:m82_40-1} shows this situation. Above the disk, at $z$ $\gg$ 
300 pc, the edge-on column density distribution shows the presence of a 
network of elongated filaments in the direction of the rarefied flow. 
Due to their high density (in spite of their low abundance), the filaments 
carry with them a large amount of metals and therefore, in this case, 
represent an efficient mechanism to transport the chemical species ejected by 
the SN explosions above the disk. 
Also, the metal abundance in the wind is large, between 5 and 10 
$Z_{\odot}$, and these values are in agreement with those obtained in previous 
studies of M82 wind \citep{strik09}. 
The launched multi-phase medium radiates energy at all  
heights (see Figure \ref{fig:m82_40-1}, bottom-right panel) and thus may be 
observed both in X and radio wavelengths.

\begin{figure}    
\begin{center}   
\psfig{figure=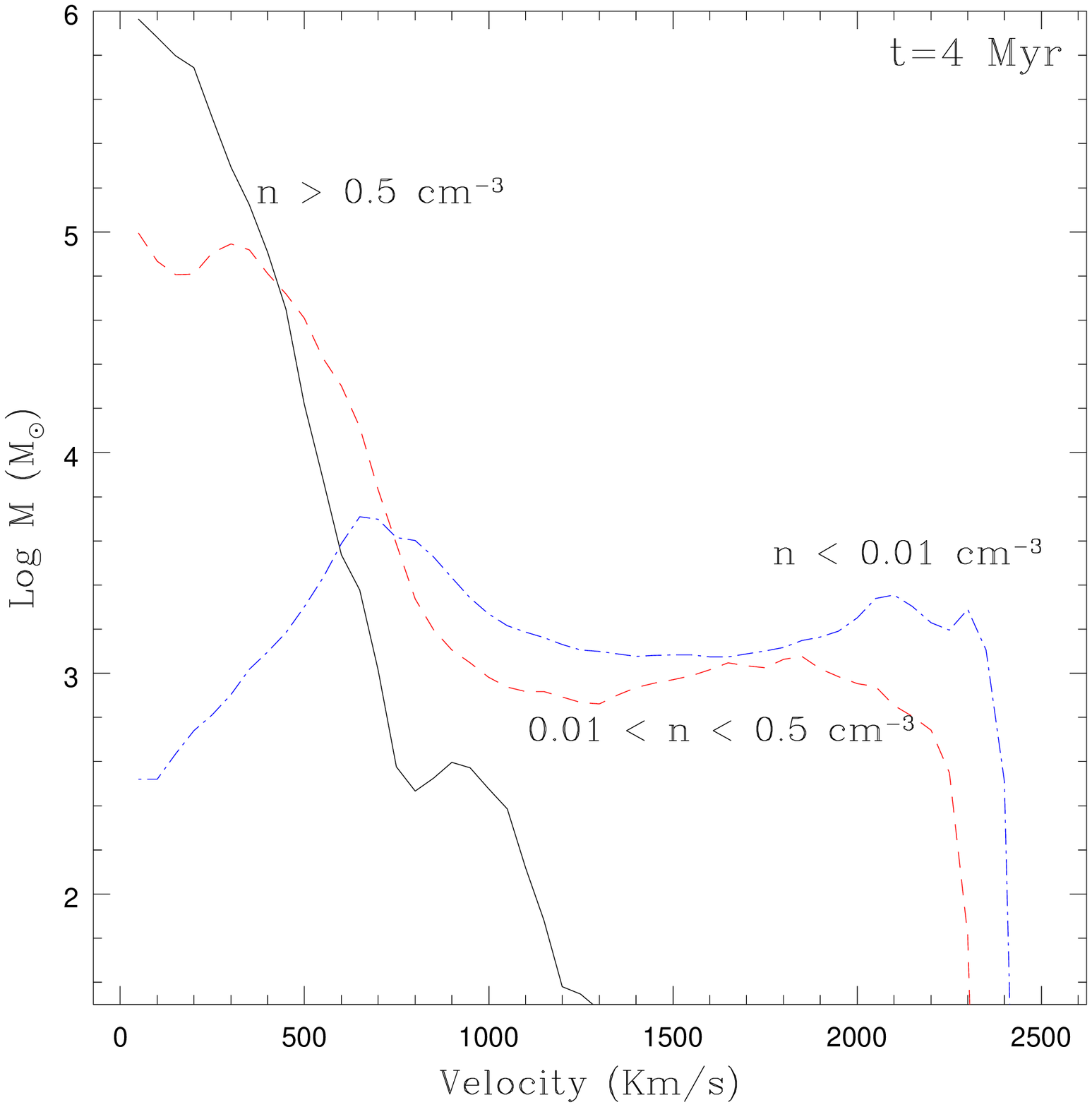,width=0.49\textwidth}    
\end{center}   
\caption{Model SC40-01: the same as in Fig. \ref{fig:V_10-10}, except that 
here we consider 40 SSCs forming over 1 Myr with a mass of 2.5 $\times 10^6$ 
M$_{\odot}$ and a SN number of 2.5 $\times 10^4$.}
\label{fig:V_40-1} 
\end{figure}

\begin{figure}    
\begin{center}   
\psfig{figure=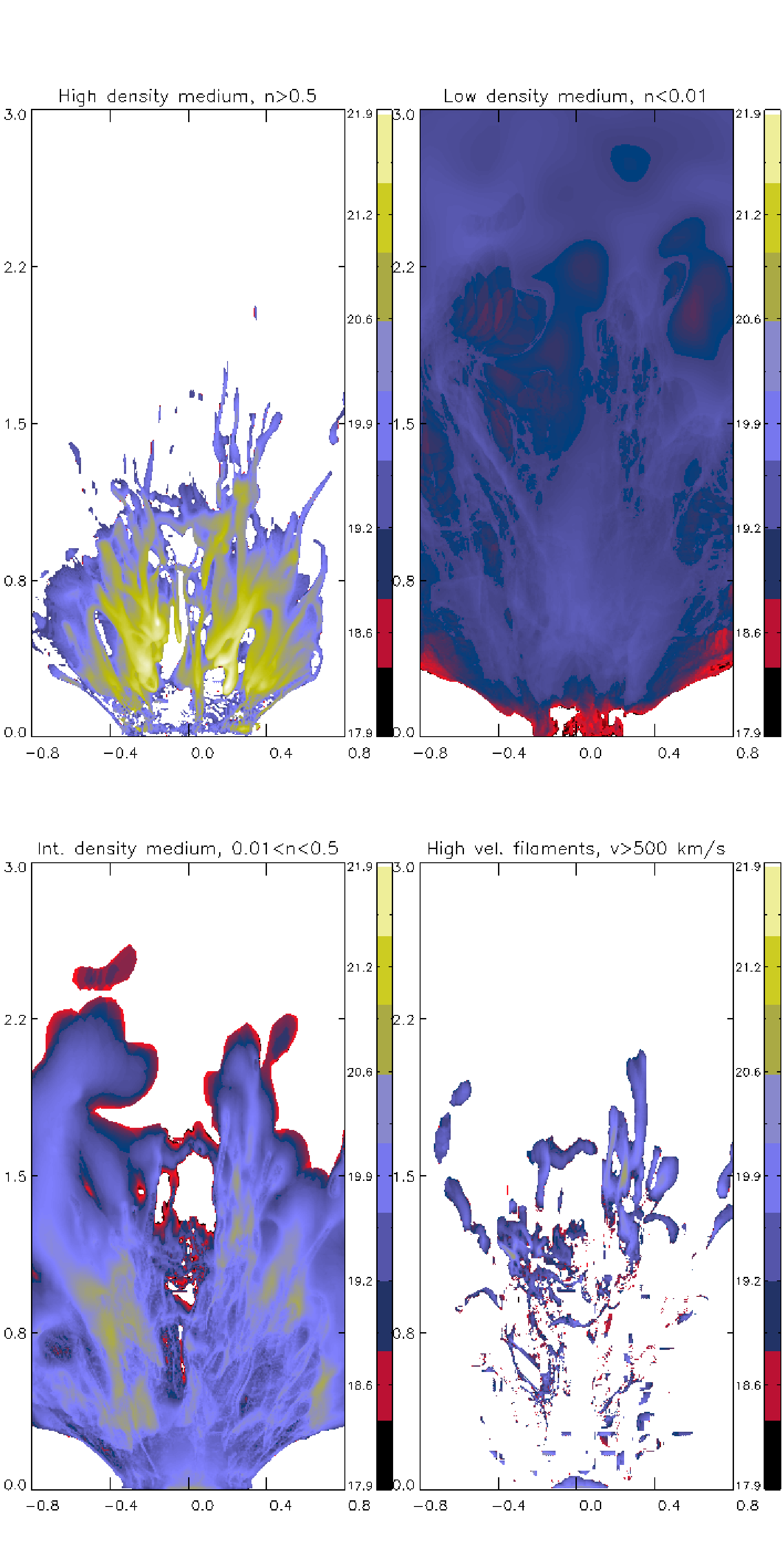,width=0.49\textwidth}    
\end{center}   
\caption{Model SC40-01: map of the gas column density for different density 
and velocity ranges, at a time of 4.5 Myr.
Top-left panel: $n \ge 0.5$ cm$^{-3}$ and $v \ge 50$ km s$^{-1}$; top-right
panel: $0.01 \le n \le 0.5$ cm$^{-3}$ and $v \ge 50$ km s$^{-1}$; bottom-left
 panel: $n \le 0.01$ cm$^{-3}$ and $v \ge 50$ km s$^{-1}$; bottom-right
panel: $n \ge 0.5$ cm$^{-3}$ and $v \ge 300$ km s$^{-1}$. The column density is
expressed in cm$^{-2}$}
\label{fig:phase_40-1} 
\end{figure}

\begin{figure}    
\begin{center}   
\psfig{figure=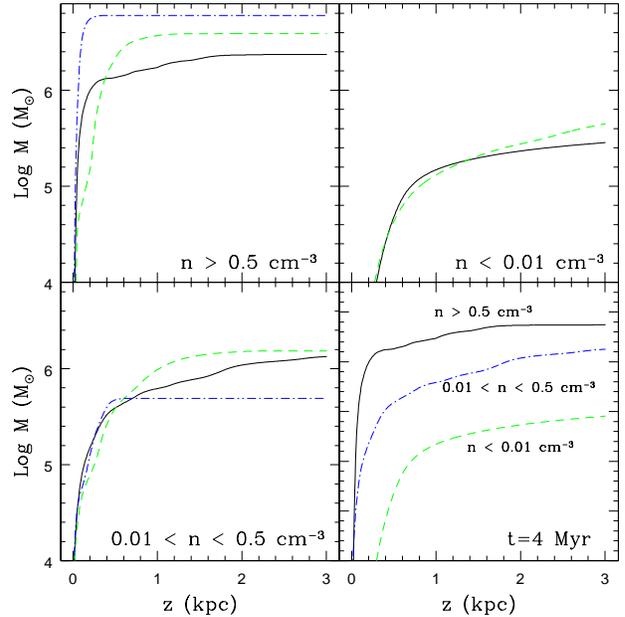,width=0.49\textwidth}    
\end{center}   
\caption{Model SC40-01: the same as in Fig. \ref{fig:fil_10-10}, except that 
here we consider 40 SSCs built up over 1 Myr with a mass of 2.5 $\times 10^6$ 
M$_{\odot}$ and a SN number of 2.5 $\times 10^4$ at a time of 1.5 Myr 
(dashed line), 3.3 Myr (dotted-dashed line) and t=4 Myr (solid line).} 
\label{fig:fil_40-1} 
\end{figure}

If we look at the velocity distribution of the different gas phases
(Figure \ref{fig:V_40-1}), we note that gas at high density with  
velocities between 150 and 800 km s$^{-1}$ has mass almost 10 times larger 
than in previous models. This means that the dense (cold)-phase
is being launched from the SB region along with the hot phase in the wind.  
This trend is confirmed by the multi-phase evolution shown in Figure
\ref{fig:phase_40-1}. The dense, intermediate and low density gas phases are 
completely superimposed, each one with its own velocity. In particular, 
the high velocity clumps and fragments are spread and embedded all over the 
wind.
The total mass of the high density phase in the flow is $\sim 4 \times 10^6$ 
M$_{\odot}$, while the gas of intermediate density has a mass of $\sim 1.5 
\times 10^6$ M$_{\odot}$, and the gas of low density has $\sim 5 \times 10^5$ 
M$_{\odot}$ (Figure \ref{fig:fil_40-1}). The high density gas in the wind is
therefore, $\sim$ 2/3 of the total GW mass, while only 8\% of the GW mass is 
due to the hot and high-velocity, low density gas.
Due to the large fraction of high density gas in the flow, most of the
injected energy is lost by radiative cooling.
After 5 Myr the total energy of the system is 2.7 $\times 10^{55}$ erg, while 
the total injected SN energy is 7.2 $\times 10^{55}$ erg. 
Since the initial energy is 8.8 $\times 10^{54}$ erg, this implies that the 
energy radiated away by radiative cooling is $\sim 5.35 \times 10^{55}$ erg, 
that is, $\sim$ 74\% of the injected energy.

\begin{figure}    
\begin{center}   
\psfig{figure=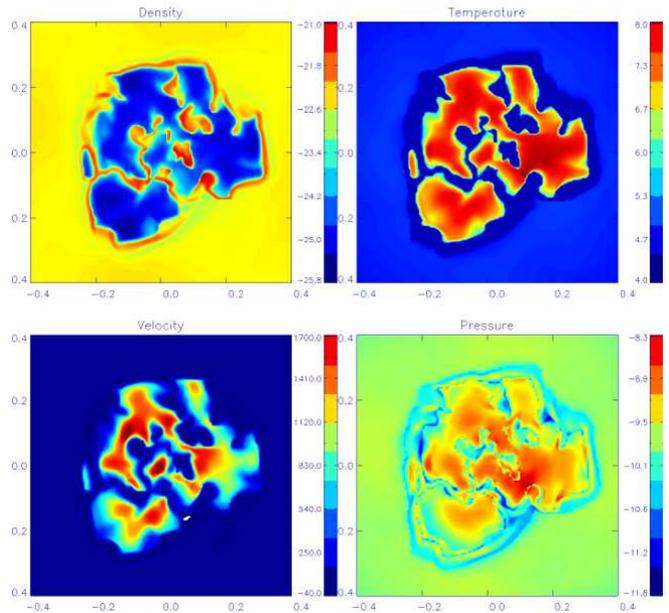,width=0.49\textwidth}    
\end{center}   
\caption{Model SC40-01: Color-scale map of the face-on M82 gas 
density (top-left panel), temperature (top-right), vertical-velocity 
(bottom-left) and pressure (bottom-right) distributions, at a height of 50 pc
above the disk and at a time t=2.5 Myr.
The density is shown in units of g cm$^{-3}$, the temperature in K, the pressure
in dyne cm$^{-2}$ and the velocity in km s$^{-1}$.}
\label{fig:m82_12-05} 
\end{figure}

As the main features of the GW in M82 seem to be similar to those
obtained in this Reference Model (Sc40-01), we can investigate in more detail 
the  physical conditions of the energized gas in this model.
Figure \ref{fig:m82_12-05} depicts the face-on gas density,
temperature, pressure and vertical-velocity distributions 
 at an earlier time  t=2.5 Myr, at a height of 50 pc above the disk 
where the coexistence of different phases is better highlighted.
A straightforward feature  in this figure is the presence of multi-phase 
structures with high temperature (T $\ge$ $10^7$ K), pressure ($p/k 
\sim 4\times 10^7)$ and density ($n \sim 5 \times 10^2$ cm$^{-3}$) 
coexisting with an underlying high velocity ($v \sim$ 2000 km s$^{-1}$), low 
density ($n \le 10^{-2}$ cm$^{-3}$) wind.
These values are similar to those inferred from recent observations
(see, e.g., \citealt{west09a,west09b}).
Moreover, looking at the vertical velocity distribution 
\footnote{We have examined several cuts along and in the  normal directions 
of the flow which are not presented here.} 
we see that the filaments produced in the core move in
the wind with a velocity between $\sim$ 100 and 800 km s$^{-1}$, while the gas 
with high thermal pressure that drives the superwind moves in different 
streams and channels with a typical velocity of $\sim$ 2000 km s$^{-1}$.
Considering that the escape velocity of the M82 wind is $\sim$ 400 km s$^{-1}$ 
(e.g., \citealt{strik09}), we can conclude that only the low density gas and a 
little fraction of the filaments may really escape from the galaxy.

\begin{figure}
\begin{center}   
\psfig{figure=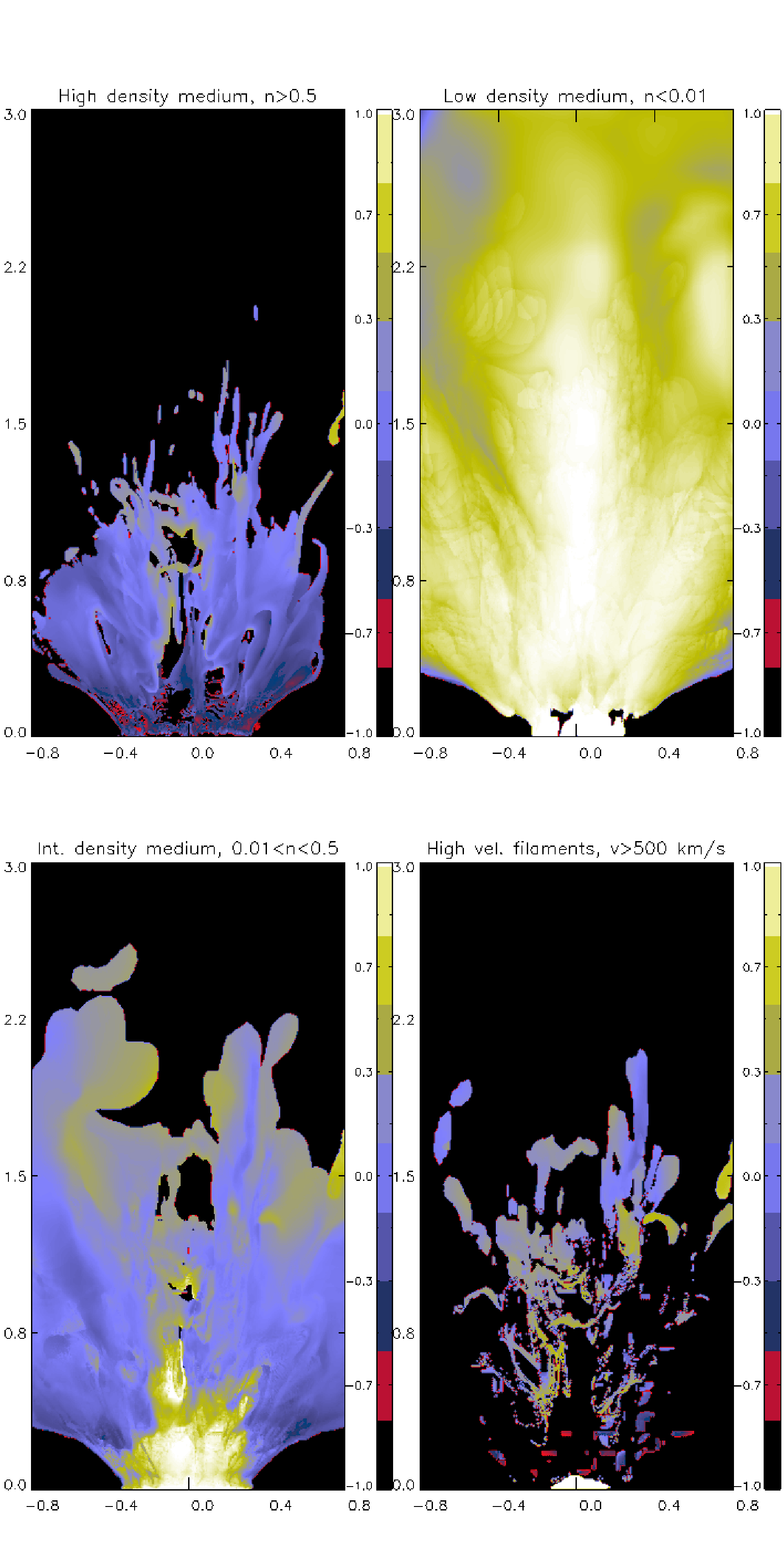,width=0.49\textwidth}    
\end{center} 
\caption{Model SC40-01: map of the average metal abundance for different 
density and velocity ranges, at a time of 4.5 Myr.
Top-left panel: $n \ge 0.5$ cm$^{-3}$ and $v \ge 50$ km s$^{-1}$; top-right
panel: $0.01 \le n \le 0.5$ cm$^{-3}$ and $v \ge 50$ km s$^{-1}$; bottom-left 
 panel: $n \le 0.01$ cm$^{-3}$ and $v \ge 50$ km s$^{-1}$; bottom-right
panel: $n \ge 0.5$ cm$^{-3}$ and $v \ge 300$ km s$^{-1}$. The abundance is
expressed in unit of solar abundance.}
\label{fig:abu_40-1} 
\end{figure}

Finally, in Figure \ref{fig:abu_40-1} we show the average abundances for
each gas-phase of the flow. The faster and smaller density phase,
with a velocity between 500 and 2000 km s$^{-1}$, is characterized by a high 
metal abundance, $\sim$ ten times larger  than the solar abundance.
On the other hand, the slower and denser phase, with a velocity between 100
and 800 km s$^{-1}$, representing $\sim$ 70\% of the total flow mass, has a 
metal abundance between 1 and 4 times the solar value.

\subsection{Model SC100-10}

The last Model (SC100-10) considers 100 SSCs built up over 10 Myr with a 
stellar mass of $10^6$ M$_{\odot}$, a SNe number of $10^4$, and a 
mean separation of $\sim$ 25 pc.
In this case, as in the models SC10-10, SC10-01 and SC-40-10, the numerical 
simulation was performed in a box domain with dimensions 1.5 $\times$ 1.5 
$\times$ 1.5 kpc.

\begin{figure}    
\begin{center}   
\psfig{figure=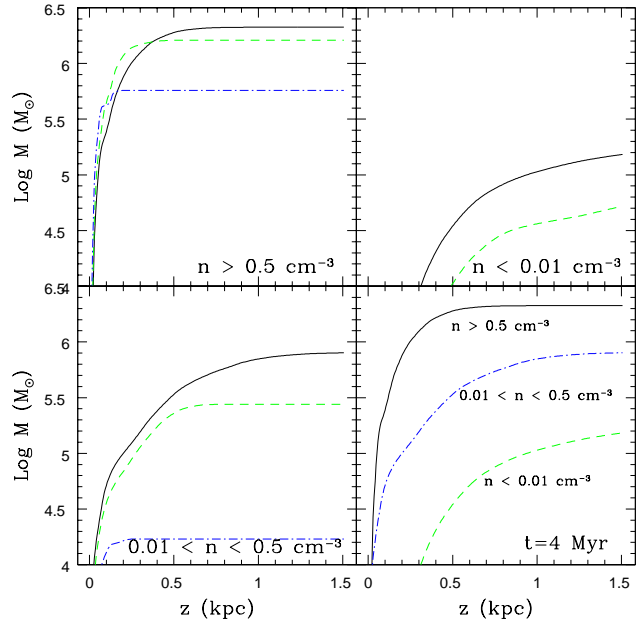,width=0.49\textwidth}    
\end{center}   
\caption{Model SC100-10: the same as in Fig. \ref{fig:fil_10-10}, except that 
here we consider 100 SSCs with a mass of $10^6$ M$_{\odot}$ and a SN number of 
$10^4$ at a time of 2.5 Myr (dashed line), 3.5 Myr (dotted-dashed 
line) and t=5 Myr (solid line).} 
\label{fig:fil_100-1} 
\end{figure}

The results are similar to those of model SC40-01.
The larger number of SSCs combined with a smaller power injected $per$ SSC 
(which are formed for longer time) generates
an ambient which is rich in filaments and with a high thermal pressure that 
accelerates the filaments and clouds to high velocities together with the hot 
wind flow. 
Also in this case
the multi-phase ambient with dense structures embedded in the hot low density 
gas provides alone 70\% of the total flow mass (Figure \ref{fig:fil_100-1}). 

The main difference with regard to the previous model (SC40-01) is the process 
of clump formation. 
In model SC40-01, after 2 Myr the filamentary structure has a 
total mass of $\sim 6 \times 10^6$ M$_{\odot}$, while after 5 Myr it decreases
to $\sim 3.5 \times 10^6$ M$_{\odot}$. In  model SC100-10, on the other hand, 
the filamentary mass $increases$ with time from $6 \times 10^5$ 
M$_{\odot}$ at $t$ = 2 Myr to $\sim 2.3 \times 10^6$ M$_{\odot}$ at 5 Myr.
Around 4 Myr the total mass of the 
filaments is about the same in both models, but in model SC40-01, a large 
amount is quickly produced and then partially ablated by the wind flow, while 
in the second case (model SC100-10), the filaments are produced more slowly 
as the superbubbles expand and collide.
This different evolution of the denser gas phase affects the radiative 
cooling too. For model SC100-10, after 5 Myr the total energy of the system is 
1.3 $\times 10^{55}$ erg, while the injected SN energy is 2.1 $\times 10^{55}$. 
This means that the energy radiated away is $\sim 1.65 \times 10^{55}$ erg, 
or $\sim$ 80\% of the total injected energy, therefore, a little larger than 
that in model SC40-01.

\section{Discussion and Conclusions}

In this study, we have investigated the formation of galactic winds (GWs) 
driven by SNe explosions, with particular focus on the M82 galaxy.

Several  driving processes of GWs have been proposed over  the last years, 
including thermal pressure due to supernova heating \citep{larson74, dek86, 
lowfer99, annFab99, strik00, mel08I, mel09II, hill12}, 
radiation pressure on dust grains \citep{martin05, murray05, nath09}, 
cosmic ray pressure \citep{breit91, everett08, uhlig12} and supersonic 
turbulence \citep{scanna10}. In most of these processes, SNe are the main 
source of energy injection which is then reprocessed. In the particular 
case of SB galaxies, where the SFR is high, we expect that the SN
energy dominates over all the other processes and therefore, in the present 
study we have only considered the role that SN shock fronts
play on the evolution of the ISM (at small scales) and on the GW evolution 
(at large scales). In the central SB of M82 there is a large number of super 
stellar clusters (SSCs) which are permeated by a complex distribution of gas 
with different densities, pressures and temperatures. As expected, most of 
our 3D radiative cooling hydrodynamical numerical simulations generated 
outflows with a complex morphology which captured the main features of an 
M82-like wind, i.e., a reach cold filamentary structure embedded in a hot low 
dense gas (see model SC-10-01, SC-40-01 and SC-100-10), starting 
right in the nuclear region and therefore in agreement with the observations
(see, e.g. \citealt{west09a}).

Although previous works have already studied the M82 wind
taking into account the effects of radiative losses, here we have performed 
a detailed analysis exploring the effects of different initial conditions in 
order to assess the connection between the characteristics of the SB 
environment and the main features of an M82-like wind.
More specifically, we tested the effects of two relevant parameters in the 
evolution of a SB GW: the total number of SSCs which are randomly distributed 
in space and time in the SB nuclear region (and their stellar mass) and the 
rate of star formation (or in other words, the build up rate of the SSCs), 
which ultimately trace  the evolution of the SNe injected power in the wind 
(see below).

In the following we discuss and summarize the main results obtained, 
particularly emphasizing the evolution and characteristics of the multi-phase
environment, the energy budget of the system, the GW metal abundance, and the 
mass and metals lost by the galaxy to the IGM.

\subsection{Multi-phase wind environment}

To study the formation and evolution of a multi-phase environment
we have considered three typical gas phases which are mainly distinguished 
by their density.
We denominated  high-density phase the gas with a number 
density $n \ge 0.5$ cm$^{-3}$; intermediate-density phase the gas with
a density between 10$^{-2}$ and 0.5 cm$^{-3}$; and  low-density 
phase the gas with a density $n \le 10^{-2}$ cm$^{-3}$.
Considering in all models the same final total injected SN luminosity  
of $10^{42}$ erg/s, but with different injection rates over the first 10 Myr
(see Figure \ref{fig:sn_inject}), we have shown that GWs are build up with 
a low-density gas component with typical velocities between 50 and 2200 km 
s$^{-1}$ and with a high-density gas component with velocities between 50 and 
800 km s$^{-1}$.

The SB region contains a total amount of gas of $\sim$ 3$\times 10^7$ 
M$_{\odot}$, but only a fraction of this gas is accelerated by
the SN explosions and driven into the wind flow, while the remaining part
is pushed to the outskirts of the SB environment (i.e. to  the external 
regions of the disk) and part remains among  the SSCs in a 
turbulent state.

Because in all models the SB region is very quickly filled 
with the superbubbles produced  by the first SSCs, the SB mass injected in 
the wind is nearly the same in all models, $\sim$ 65\% of the total mass at the 
beginning of the SN activity, i.e. $\sim$ 2 Myr, settling then
to about 75\% at the end of the simulations, at 5 Myr. 
However, depending on the model,  the diffuse gas may be either converted in 
dense filaments and clumps due to the compression and radiative cooling of 
the swept material by the SN shock waves, or just ejected
from the SB region in a hot, low density wind surrounded by a thin, dense 
shell. 
This is consistent with the results found by \citet{tenorio03} and 
\citet{rod08a,rod08b}.
In both cases a GW develops, but with distinct morphology.
In the first case we have "filamentary-like" GWs, as observed in M82 and also 
in models SC10-01, SC40-01 and SC100-10, while in the second case we 
have "supperbubble-like" GWs where the superbubble expands until its external 
super-shell fragments allowing the hot gas to flow into the IGM
and where the only structures are those generated by the fragmentation of the 
super-shell, as in models SC10-10 and SC40-10.
We note that, regardless of the total power injected which is the same for all 
models ($10^{42}$ erg/s), the spatial and temporal distribution of the SSCs 
within the SB region and thus  the first million years of SNe 
activity (Figure \ref{fig:sn_inject}), are determinant factors to allow 
the formation of a rich filamentary structure in the wind. 
If the SSCs form fast enough and if they are in a sufficiently large number 
then, several SN shells evolve overall the region and can interact nearly 
simultaneously, building up a rich filamentary structure which is dragged by 
the wind.

Considering the discussion above, we may try to establish the physical 
conditions which are necessary to form a multi-phase ISM and an M82-like GW. 
We have found that a larger amount of clumps and filaments is mainly generated 
during the first interactions of the SNe superbubbles when their volume 
filling factor is $ff \le 1$  and a large fraction of unperturbed gas 
is still in the nuclear region. Therefore, it is the number of active SSCs 
before $ff$ becomes $ff$=1 that determines whether the wind will be  a 
``filamentary'' or  a ``superbubble-like'' wind.
Our  results indicate  that $ff$ becomes equal to unity at $t$=1 Myr for  model
SC10-10, at $t$=0.6 Myr for  model SC10-01, at $t$=1.3 Myr for  model
SC-40-10, at $t$=0.65 Myr for  model SC40-01, and at $t$=1.9 Myr for
model SC-100-10. At these times, the number of active SSCs is 3 
for SC10-10, 7 for SC10-01, 4 for SC40-10, 26 for SC40-01 and 23 for SC100-10. 
This means that the larger the number of SSCs formed before $ff$ becomes 
$ff=1$, the more complex and rich the wind filamentary structure will be. 
When the number of active SSCs is $\ge$ 7-10 and $ff\le$1 (which corresponds 
to a mean separation between simultaneously active SSCs $\le$ 80 pc), 
a multiphase environment is built up in the central region of the galaxy 
favouring the formation of a filamentary-like wind, as in M82.

The resulting filaments have extensions between $\sim$ 50 and 500 pc, 
thickness of $\sim$ 30 pc, densities between 1 and 10 cm$^{-3}$, temperatures 
of about 10$^4$ K, pressures between 10$^{-11}$ and 10$^{-12}$ dyne cm$^{-2}$ and 
velocities between $\sim$ 100 and 800 km/s.
Outside of the disk, because the wind has a velocity $\sim $2200 km/s, 
they are impacted by the hot gas with  relative velocities of 1500-1800 km/s
and therefore, the 
temperature may increase up to 10$^7$ K.
Just a little fraction of these filaments escapes from the galaxy and, 
based on the results of this work, this corresponds to only $\sim$ 10\% 
of the gas in the high density phase, that is, about 5$\times 10^5$ 
M$_{\odot}$. This result suggests that although the wind can be
efficient to inject energy into the halo, it seems to be less efficient to 
inject disc mass (see also Section 4.4).

We can conclude that the environmental conditions at the base will determine
the main large-scale features of the GW morphology and particularly, whether
filaments and a multi-phase environment will develop or not in the wind.
If a large number of SSCs form before the SNe driven hot superbubbles fill out 
the entire nuclear volume of the galaxy, then an M82-like wind, with a rich 
filamentary structure, develops. On the other hand, if only a few, highly 
massive SSCs form in the SB region then the SNe will drive a superbubble-like 
wind with no filaments and with all the  phases of the gas (high, intermediate 
and low density) well separated from each other.

The results above also suggest that the features of the wind mainly 
depend on the recent star formation history of the starburst rather than all 
earlier starburst episodes which occurred  before $\sim$ 50 Myr. 
In fact, although many observations indicate that intense episodes of star 
formation occurred in M82 also about 100 Myr ago (see, e.g., \citealt{deG01}), 
the simulations indicate that the present wind energy and morphology are 
essentially due to the recent activity occurring along few Myr 
of the younger SSCs lifetime. 
Remnants of fossil star formation activity had been swept long ago by the 
strong power driven by previous SNe explosions and are possibly in the outer 
regions of the wind or have already evaporated. For this reason, it is 
sufficient to consider only the last $\sim$10 Myrs of the starburst lifetime 
in order to assess the main features of the large scale wind structure in the 
galaxy surrounds, as in the models discussed here.

\subsection{Energy budget}

\begin{figure}    
\begin{center}   
\psfig{figure=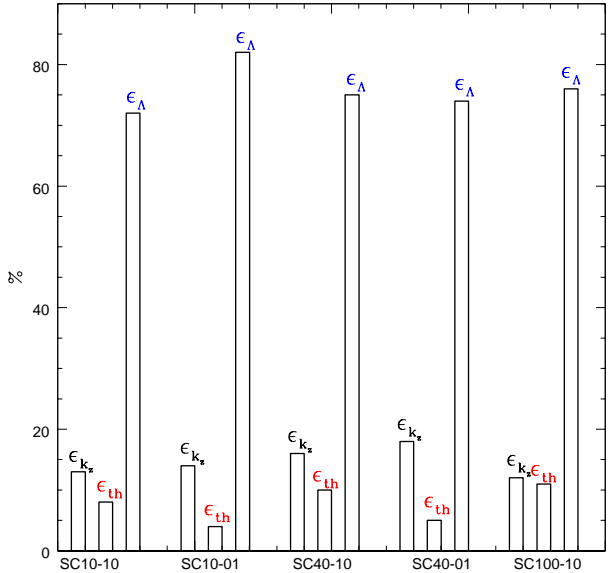,width=0.49\textwidth}    
\end{center}   
\caption{SN energy distribution for the different models studied here after 
a time of 5 Myr. $\epsilon_{k_z}$: fraction of the total SNe energy stored in 
the gas as vertical (or poloidal) kinetic energy; $\epsilon_{th}$:
fraction of the total SNe energy stored in the gas as thermal energy; 
$\epsilon_{\Lambda}$: fraction of the total SNe energy lost by radiative 
cooling.}
\label{fig:ene}
\end{figure}

The energy injected by the explosions of SNe from a single SSC has a heating
efficiency $HE \sim$ 100\% (where, as remarked in Section 2.2, HE corresponds 
to the ratio between the kinetic plus thermal energies stored in the gas and 
the energy injected by the SN explosions, \citealt{mel04}). 
This is  because the radiative losses are 
important only if t$_{cool}$ $\le$ t$_{int}$ (where t$_{cool}$ and t$_{int}$ are 
the gas radiative cooling time and the interaction time of the supernova 
remnants, respectively) and in the case of single SSC the SNe occur so close 
in space and time that this  relation is almost never fulfilled.
But what fraction of the total energy of the superbubbles goes to drive bulk 
gas motions and generate a large scale wind?
In all the models, we have found that the energy lost by radiative cooling is
proportional to the ratio between the mass of the high-density gas component 
and the total mass of the GW. Since this value ranges between 70\% and 82\%, 
we verify that only a fraction between 20\% and 30\% of the SN energy goes 
into the kinetic energy of the gas to drive the GW (Figure \ref{fig:ene}). 
Therefore, in contrast to the results obtained for the formation of a 
multi-phase environment, the fraction of energy available for the build up 
of the GW is about the same for all the models independent on the temporal 
and space distribution of the SSCs  and on the
SNe power injection during the first Myr.

\begin{figure}
\begin{center}
\psfig{figure=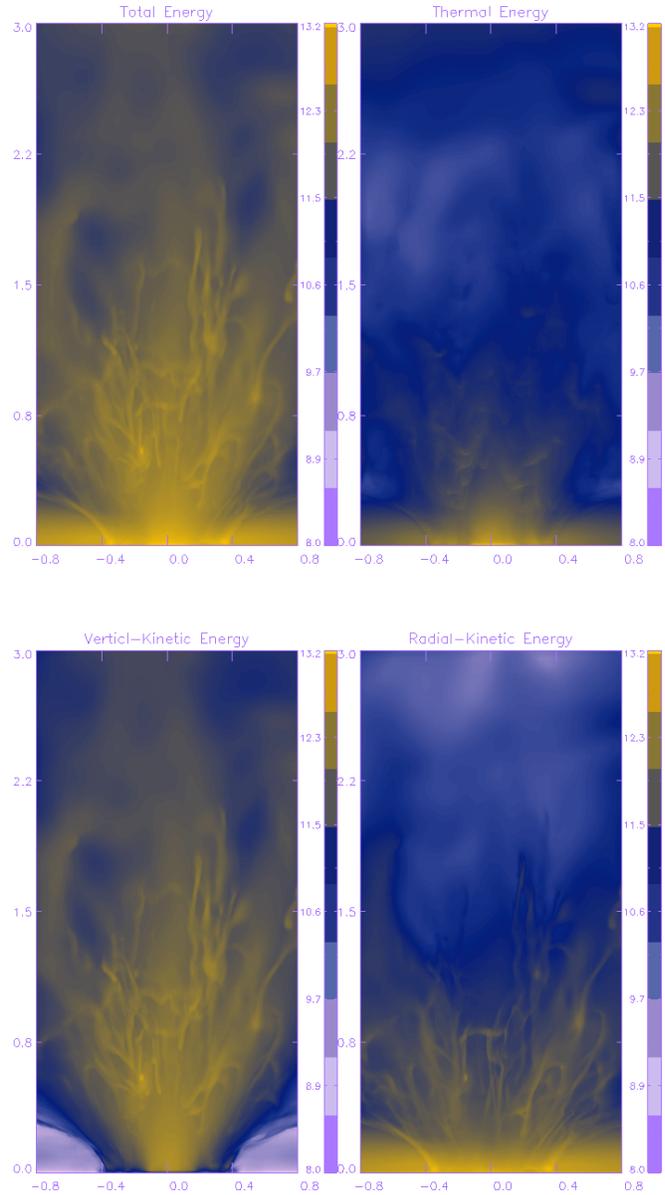,width=0.49\textwidth}    
\end{center}   
\caption{Model SC40-01: map of the energy "column density" (i.e., the energy 
density integrated in the transverse direction to the wind) at a time 3.5 Myr.
Top-left panel: total energy; top-right panel: thermal energy; bottom-left 
panel: vertical or poloidal-kinetic energy; bottom-right panel: radial-kinetic 
energy. The energy column density is expressed in erg cm$^{-2}$.}
\label{fig:ene1} 
\end{figure}

The injected energy that is not radiated away provides most of the 
kinetic energy to the gas in the high-density phase (filaments and clouds) 
and the thermal energy to the gas in the intermediate and low-density phases. 
These energies are spatially distributed as depicted in Figure \ref{fig:ene1} 
for model SC40-01. 

These results suggest that the average temperature of the hot gas will be 
given by:
\begin{equation}
T \sim \frac{2}{3} \frac{\epsilon_{th} E_{SN}}{n_{i+l} k}
\end{equation} 
where $\epsilon_{th}$ is the fraction of the total SNe energy stored in the gas 
as thermal energy and $n_{i+l}$ is the total number density of the 
intermediate and low-density gas components.

Similarly, the average velocity of the filaments will be given by:
\begin{equation}
v_f \sim \sqrt{\frac{2 \epsilon_{k} E_{SN}}{m_{h}}}
\end{equation}
where $\epsilon_{k}$ is the fraction of the SNe total energy stored in the gas 
as kinetic energy, and $m_{h}$ is the total mass of the  high-density gas.

With the results of the Reference Model (SC40-01) presented in 
Figure \ref{fig:ene}, we obtain an average temperature for the low-density gas  
$T \sim 10^7$ K and an average velocity for the  high-density gas 
$v_f \sim 500$ km s$^{-1}$, where we have considered  $\epsilon_{th}$ $\sim$ 5\%
and $\epsilon_{k}$ $\sim$ 18\% (see Figure 20).
This is in agreement with \citet{chev85}
who predicted that for a thermalization efficiency of about 10$\%$ the wind 
fluid should have a temperature in a range between 10$^7$ and 10$^8$ K even if 
it had been slightly mass loaded with cold ambient gas. 
We note that although the wind model of \citet{chev85} is adiabatic, 
these authors introduced two parameters in their model ($\alpha$ and $\beta$) 
in order to incorporate the effects of radiative cooling both from the SN 
ejecta and dense gas, so that a comparison of their model with the present 
ones is appropriate.
However, as pointed out by \citet{strik09}, such temperatures are hotter than 
the usually observed warm ionized, soft X-ray emitting plasma in GWs. 
Thus hard X-ray observations are still required in order to confirm or not
the predictions of the models.

Based on the results above of the energy distribution during the wind 
evolution, we can conclude that an M82-like GW could be generated in two steps. 
The first one can be characterized by the  formation of supperbubbles due to 
SNe activity in each SSC with a SN heating efficiency HE $\sim 100\%$ 
\citep{mel04}. 
The second step is characterized by large scale interactions of the 
superbubble shells with a HE (including the thermal and the kinetic energy 
contributions) of $\sim$ 20\% only. In this scenario, about 80\% of 
the SN energy is radiated away mainly by the  denser structures 
embedded in the GW.

\subsection{Chemical evolution}

Our simulations show that a large fraction of the metals injected into the 
ISM by the SNe are in the filaments located near the galactic disc
(see the upper-right panel of Figure \ref{fig:m82_40-1}).
In fact, the metal abundance in this high-density gas component is small, but 
the total amount of metals it contains is similar to the metal mass contents 
in the low-density component.

Regardless of the model, the high-density, high-velocity
structures have a maximum abundance close to the solar one, while the 
low-density phase is characterized by a maximum abundance of $\sim$ 10 
$Z_{\odot}$. Examining the maps of Figure \ref{fig:abu_40-1} for our Reference 
Model SC40-01, we can determine an average abundance,
$\bar{Z}$, for each gas phase. For the high-density phase we obtain
$\bar{Z_h} \sim 0.5$ $Z_{\odot}$, for the intermediate-density phase 
$\bar{Z_i} \sim 1.2$ $Z_{\odot}$, while for the low-density phase we have
$\bar{Z_l} \sim 4.5$ $Z_{\odot}$.
Each gas phase regulates the chemical evolution of the different 
parts of the system.
The high and intermediate-density phases, with their filaments and clumps, are
unable to escape from the galaxy and are thus responsible for retaining in
it an important fraction of the metals ejected by the SN explosions.
In fact the total amount of ejected 
metals after 8 Myr is $\sim$ 2$\times 10^5$ M$_{\odot}$, but the amount which 
escapes from our physical domain is only $\sim$ 8$\times 10^4$ M$_{\odot}$ and
the remaining portion is retained in and around the galactic disk.
The total mass of metals in the high and intermediate-density phases
are, respectively, $M_{h}\times \bar{Z_h}$ and $M_{i}\times \bar{Z_i}$, that
is, $\sim$ 4$\times 10^4$ M$_{\odot}$ of metals for each phase, which represents
$\sim$ 2/3 of the metals stored in the galaxy.
The remaining 1/3 of the chemical species is in the low-density phase and 
is therefore, the only fraction of metals that is transported to outside of 
the galaxy to enrich the IGM.

Based on the results above, we can conclude that the SNe explosions in an 
M82-like SB galaxy change significantly the metallicity only of the 
low-density wind component.  
In our simulations this component has a resulting metal abundance 
between 5 and 10 $Z_{\odot}$. The injected metals by the SNe do not  
affect significantly the metallicity of the high and intermediate
density phases of the wind, a result which is in contrast to what occurs, 
for instance, in the clouds built up in galactic fountains 
\citep{mel08I, mel09II}.

\subsection{Mass loss}

Our results suggest that during the early phases of 
expansion, the GW transports to  above the disk a large amount of 
energy, momentum and gas, but at the same time the high-density component (of 
clouds and filaments) is able to reach only intermediate altitudes and does not 
escape from the system.
As a consequence, no significant amount of gas mass is lost in the IGM, 
except for an important fraction of metals and SN ejecta that flows along 
with the high velocity-low density gas in the GW.
In our  Reference Model (SC40-01), for instance, the GW reaches a nearly 
steady state 
evolution after 4-5 Myr and until  that moment the mass lost by the system is 
about 0.6 M$_{\odot}$ yr$^{-1}$ only. In fact, as stressed in the previous 
Section (Sec. 4.3), the only gas phase which is able to flow freely into the 
IGM is the low-density phase. However, due to its very low-density, the total 
gas mass in this phase represents only  5\% of the total mass of the GW 
(see Figure \ref{fig:fil_40-1}). We note that this value is only slightly 
smaller than that indicated by \citet{strik09}.

Based on the results above, we can conclude that the mass evolution of the 
galaxy is not substantially affected by the SB events observed in the core of 
M82.

\section{Summary}

The main conclusions of this work can be summarized as follows:

\begin{enumerate}

\item The environmental conditions at the base will determine
the main large-scale features of the GW morphology.
If a sufficiently large number of super stellar clusters (SSCs) is build up 
before the SNe driven hot superbubbles fill out the entire nuclear volume of 
the galaxy, then an M82-like wind, with a rich filamentary structure, develops.
In quantitative terms, a multiphase wind is build up, as in M82, if the number 
of SSCs is $\ge$ 7-10 while  $ff$ is still $ff\le 1$.

\item About 80\% of the SN energy is radiated away mainly by the high density 
structures which are formed from the fragmentation of the superbubble shells 
(see e.g., Model SC40-01).

\item The SNe explosions in an M82-like galaxy change significantly the 
metallicity only of the low-density wind component.  
In our simulations this component has a resulting metal abundance 
between 5 and 10 $Z_{\odot}$, while the metallicity of the high and intermediate
density components is around the solar abundance (i.e., $\sim 0.5$ to 1 
$Z_{\odot}$).
Also according to our results, about 1/3 of the total mass in metals is in the 
low-density component and therefore, this is the only fraction that can be 
removed to the IGM.

\item During the early phases of 
expansion, the GW transports to outside the disk large amounts of 
energy, momentum and gas, but the more massive high-density component 
(which is composed of clouds and filaments) is able to reach only intermediate 
altitudes (typically smaller than 1.5 kpc) and therefore, does not 
escape from the system. In consequence,
no significant amounts of gas mass are lost to the IGM and the mass evolution 
of the galaxy is not  much affected either by the starburst events occurring 
in the nuclear region.

\end{enumerate}

\section*{Acknowledgments}
C.M. acknowledges support from a post-doctoral fellowship from the Brazilian 
agency FAPESP (grant 2011/22078-6); E.M. G.D.P. acknowledges partial support 
from FAPESP (2006/50654-3) and from CNPq (grant no. 300083/94-7).

\bibliographystyle{mn2e} \bibliography{msREV13_ref}

\label{lastpage}
\end{document}